\documentclass{amsart}

\usepackage{amsmath}
\usepackage{bm}
\usepackage[margin=1in]{geometry}
\usepackage{amsmath,amsfonts}
\usepackage{amsthm}
\usepackage{natbib}
\usepackage{bbm}
\usepackage{nicematrix}
\usepackage{graphicx}
\usepackage{subcaption}
\usepackage{url}
\usepackage{tikz}
\usetikzlibrary{arrows,positioning,shapes.multipart}
\usepackage{amsaddr}
\usepackage{setspace}
\usepackage[mathlines]{lineno}
\usepackage{appendix}

\newtheorem{theorem}{Theorem}
\newtheorem{corollary}{Corollary}
\newtheorem{remark}{Remark}
\newtheorem{lemma}{Lemma}
\newtheorem{proposition}{Proposition}

\title[Stability of difference equations with maturation delays]{Stability of difference equations with interspecific density dependence, competition, and maturation delays}
\author{Geoffrey R. Hosack, Maud El-Hachem, Nicholas J. Beeton}
\address{CSIRO Data61, 3 Castray Esplanade, Hobart, Tasmania 7001}

\begin{document}

%\linenumbers

\maketitle

\begin{abstract}
A general system of difference equations is presented for multispecies communities with density dependent population growth and delayed maturity. Interspecific competition, mutualism, predation, commensalism, and amensalism are accommodated. A sufficient condition for the local asymptotic stability of a coexistence equilibrium in this system is then proven.  Using this system, the generalisation of the Beverton-Holt and Leslie-Gower models of competition to multispecies systems with possible maturation delays is presented and shown to yield interesting stability properties. The stability of coexistence depends on the relative abundances of the species at the unique interior equilibrium. A sufficient condition for local stability is derived that only requires intraspecific competition to outweigh interspecific competition. The condition does not depend on maturation delays. The derived stability properties are used to develop a novel estimation approach for the coefficients of interspecific competition. This approach finds an optimal configuration given two conjectures. First, coexisting species strive to outcompete competitors. Second, persisting species are more likely in stable systems with strong dampening of perturbations and high ecological resilience. The optimal solution is compared to estimates of niche overlap using an empirical example of malaria mosquito vectors with delayed maturity in the \textit{Anopheles gambiae} sensu lato species complex.
\end{abstract}

\section{Introduction}

\setdisplayskipstretch{3}

Delayed maturity, where juveniles have zero fecundity before maturing into reproductive adults, and interspecific density dependence caused by the fundamental ecological interactions of competition, mutualism, predation, commensalism or amensalism are two important factors that determine the stability of ecosystems and ecological communities \citep{May1974}. Sufficient conditions for local asymptotic stability of equilibria are therefore derived for an $m$-dimensional vector of species abundances $\bm{x}(t) = [x_1(t), \ldots, x_m(t)]^\top$  in the system of delay difference equations \citep{Fisher1984sys},
\begin{equation}
	x_i(t + 1) = \sigma_{\delta_i \mid i} x_i(t) + F_i(\bm{x}(t - \delta_i))x_i(t - \delta_i), \quad i \in \mathcal{M} = \{1, \ldots, m\}, \quad t \in \mathbb{Z}.\label{eq:DDeq}
\end{equation}
In Eq. \eqref{eq:DDeq}, each species $i \in \mathcal{M}$ attains adulthood and reproductive maturity at age $\delta_i \in \mathbb{N}_0 = \{0, 1, 2, \ldots\}$. The adults of species $i$ with age greater or equal to $\delta_i$ survive to the next time step with age-class specific probability $\sigma_{\delta_i \mid i} \in (0, 1)$ and complementary probability of adult mortality, $z_i = 1 - \sigma_{\delta_i \mid i}$. Adults reproduce with per capita population growth function $F_i$ that accounts for immature survival, the delayed maturity of species $i$, and both intraspecific and interspecific density dependence. 

The single species case ($m = 1$) of Eq. \eqref{eq:DDeq} corresponds to the well-known model proposed by \citet{Clark1976}. A short summary of findings for this important equation are as follows. \citet{Clark1976} presents a sufficient condition for local stability at an equilibrium $x_1^\ast$ that does not depend on the maturation delay,
\begin{equation}
	\left| \left. \frac{d}{dx_1(t)}\, x_1(t) F_1\left(x_1(t)\right) \right|_{x_1(t) = x_1^\ast}  \right| < z_1 = 1 - \sigma_{\delta_1 \mid 1}. \label{eq:Clark}
\end{equation}
However, necessary and sufficient conditions do depend on the delay and also require numerical calculation outside special cases \citep{Clark1976}.  The necessary and sufficient local stability conditions are shown by \citet{Kuruklis1994} and \citet{Papanicolaou1996} to generalise even for real-valued $\sigma_{\delta_1 \mid 1} \in \mathbb{R}$. Local stability for the single species model in certain cases also implies global stability, but not generally \citep{Lopez2016}. \citet{Fisher1984} propose sufficient conditions for global stability by construction of a Lyapunov function using results from \citet{LaSalle1976}. These sufficient conditions also do not depend on the delay.  Elaborations based on \cite{Clark1976} are  not considered further here, for example, \citet{Goh1980} and \citet{Streipert2023} generalise to density dependent survival functions. See \citet{Liz2020} for a recent review and further analysis  of single species delayed difference population models based on Eq. \eqref{eq:DDeq} with $m = 1$.

The multispecies delayed system with $m > 1$ in Eq. \eqref{eq:DDeq} has received far less attention compared to the single species case. Exceptionally, \citet{Fisher1984sys} shows that the local and global stability properties are deduced from the corresponding non-delayed system defined by
\begin{equation}
	x_i(t + 1) = z_i^{-1}F_i(\bm{x}(t)) x_i(t), \quad z_i = 1 - \sigma_{\delta_i \mid i} \in (0, 1), \quad i \in \mathcal{M}.\label{eq:nondelayed_system}
\end{equation}
In the important case of interspecific competition in non-delayed systems, \citet{Sacker2011} states that consideration of multispecies discrete-time systems ``inevitably'' leads to consideration of the multispecies extension of the Leslie-Gower model \citep{Leslie1958} equivalent to the non-delayed system with 
\begin{equation}
	x_i(t + 1) = F_i(\bm{x}(t)) x_i(t) = \frac{\lambda_i x_i(t)}{1 + \sum_{k \in \mathcal{M}} A_{ik} x_k(t)}, \quad i \in \mathcal{M},\label{eq:MLG}
\end{equation}
for $m$-dimensional positive fecundity rate vector $\bm{\lambda} > \bm{0}$ and $m \times m$ nonnegative matrix $\bm{A} \geq \bm{0}$ composed of nonnegative entries $A_{ij} \geq 0$ for all $i, j \in \{1, \ldots, m\}$. The interspecific competitive effect of species $j$ on the population growth rate of species $i \neq j$ is defined by $A_{ij}$, and intraspecific competition is defined by $A_{ii}$. \citet{Sacker2011} provides sufficient conditions for global asymptotic stability of the unique positive equilibrium in the non-delayed model that depends on ``sufficiently small'' interspecific competition defined by $\bm{A}$, the entry-wise inequality $\bm{\lambda} > \bm{1}$, and the condition
\begin{equation}
	\lambda_i - 1 \geq \sum_{j \in \mathcal{M}\backslash i} \lambda_j \frac{A_{ij}}{A_{jj}}, \quad i \in \mathcal{M}.\label{eq:Scond}
\end{equation}
Using the results of \citet{Fisher1984sys}, global stability of the interior equilibrium in the non-delayed Eq. \eqref{eq:MLG} by Eq. \eqref{eq:nondelayed_system} induces global stability for the identical interior equilibrium of the corresponding delayed system of Eq. \eqref{eq:DDeq}. The condition of global stability for the delayed system thus depends not only on the density-independent fecundity rates via Eq. \eqref{eq:Scond} but also the adult survival probabilities via Eq. \eqref{eq:nondelayed_system}.

In Section \ref{sec:model}, sufficient conditions are given for local stability in the general multispecies delayed model of Eq. \eqref{eq:DDeq}. The sufficient conditions for stability derived by linearisation of an equivalent age-specific version provide the multispecies analogue of the sufficient condition of Eq. \eqref{eq:Clark} for the single species case. Constraints on the eigenvalues of its linearised approximation near an equilibrium are obtained using a geometric argument by Ger{\u s}gorin's Theorem \citep{Gersgorin1931} as extended by \citet{Brualdi1982} using directed graphs. The result is applied to a competitive delay multispecies (CDM) model. The CDM model extends the multispecies Leslie-Gower model of competition of Eq. \eqref{eq:MLG} to allow for potential delayed maturity and age-specific survival probabilities (Section \ref{sec:multi_comp}). Sufficiency conditions are established such that a positive coexistence equilibrium, if it exists, is both unique and locally asymptotically stable if interspecific competition does not outweigh intraspecific competition. Importantly, these conditions hold even if the exact values of the delays, density independent fecundity rates and age-specific survival probabilities in the CDM model are unknown or uncertain.

It is also shown that the local asymptotic stability of the interior equilibrium for the CDM model depends on the relative abundances, that is, the species composition or fractional contributions of each species to the total equilibrium abundance. This result is used in Section \ref{sec:estimation} to develop a method to find an optimal configuration of the parameters $\bm{A}$ given the assumptions that i) species are more likely to persist in a system with greater resilience and faster return times to a stable equilibrium after disturbances, and ii) species seek to increase competitive advantage over other species. The estimate requires knowledge of fecundity, density independent survival probability and an observed equilibrium in which species coexist. The approach does not require specific information about the carrying capacities of the species, but instead depends only on estimates of relative abundance that are often more readily available from field data than absolute or true abundance \citep{Sinka2016}. The development extends ideas that relate relative abundance to measures of niche overlap and parameterisation of the competition coefficients in the Lotka-Volterra competitive system of ordinary differential equations \citep{Levins1968, May1975}. The previous and novel procedures are compared using empirical estimates provided by \citet{Pombi2017} for a system of three competing species of \textit{Anopheles gambiae} sensu lato mosquitoes with delayed maturity that vector human malaria.

\section{Interspecific Density Dependence with Delayed Maturity}\label{sec:model}

\setdisplayskipstretch{1.5}

An equivalence exists between a delayed difference equation system and an expanded non-delayed system with explicit age structure, which has been applied to both single species \citep{Fisher1984, Streipert2023} and multiple species \citep{Fisher1984sys} examples. It can be shown that Eq. \eqref{eq:DDeq} is indeed a collapsed version of an expanded age-specific model with age-specific immature survival probabilities. To see this, let $a \in \{0, \ldots, \delta_i - 1\}$ denote the age-classes of species $i$ with immature age-specific survival probabilities $\sigma_{a \mid i} \in  (0, 1]$. Define the vector-valued per capita growth function $\bm{G}$ adjusted for the explicit inclusion of immature survival probabilities such that
\begin{equation}
	G_i(\bm{x})\prod_{a = 0}^{\delta_i - 1} \sigma_{a \mid i} = F_i(\bm{x}), \quad i \in \mathcal{M}.\label{eq:FG}
\end{equation} 
If $\delta_i = 0$ then the empty product $\prod_{a = 0}^{\delta_i - 1} \sigma_{a \mid i}$ is defined as the multiplicative identity with value one. Define the vectors of age-specific abundances with size $(\delta_i + 1)$ for each species at time $t$ by
\begin{equation*}
	\bm{y}_i(t) = [y_{0 \mid i}(t), \ldots, y_{\delta_i \mid i}(t)]^\top,
\end{equation*}
where $y_{a \mid i}(t)$ is the abundance of age class $a \in \{0, \ldots, \delta_i\}$ for species $i$ at time $t$. Define also the vector $\bm{y}(t) = [\bm{y}_1(t)^\top, \ldots, \bm{y}_m(t)^\top]^\top$ of size  $(m + \sum_{i \in \mathcal{M}} \delta_i)$ of all age classes for all species at time $t$. Let $\bm{h}$ denote the vectorised function such that $\bm{y}(t + 1) = \bm{h}(\bm{y}(t))$, where $y_{a \mid i}(t + 1) = h_{a \mid i}(\bm{y}(t))$ for $a = 1, \ldots, \delta_i$ and $i \in \mathcal{M}$. Let the $m$-dimensional vector
\begin{equation*}
	\bm{y}_{\bm{\delta} \mid \mathcal{M}}(t) = [y_{\delta_1 \mid 1}(t), \ldots, y_{\delta_m \mid m}(t)]^\top = \bm{x}(t)
\end{equation*}
denote the adult abundances of the $m$ species at time $t$, where $\bm{\delta} = [\delta_1, \ldots, \delta_m]^\top$.  The system of recurrence equations for $\bm{y}(t)$ composed of ages $a \in \{0, \ldots, \delta_i\}$ for each species $i \in \mathcal{M}$ at times $t \in \mathbb{Z}$ is then defined by
\begin{equation}
	y_{a \mid i}(t + 1) = h_{a \mid i}\left(\bm{y}(t)\right) =
	\begin{cases}
		G_i\left(\bm{y}_{\bm{\delta} \mid \mathcal{M}}(t)\right) y_{\delta_i \mid i}(t)  & 
		\textrm{if } a = 0  \textrm{ and } \delta_i > 0\\
		\sigma_{a - 1 \mid i}y_{a - 1 \mid i}(t) & \textrm{if } a \in \{1, \ldots, \delta_i - 1\} \textrm{ and } \delta_i > 0 \\
		\sigma_{\delta_i - 1 \mid i}y_{\delta_i - 1 \mid i}(t) + \sigma_{\delta_i \mid i}\ y_{\delta_i \mid i}(t) & \textrm{if } a = \delta_i  \textrm{ and } \delta_i > 0 \\
		\sigma_{0 \mid i}y_{0 \mid i}(t) + G_i\left(\bm{y}_{\bm{\delta} \mid \mathcal{M}}(t)\right) y_{0\mid i}(t) & \textrm{if } a  = \delta_i = 0.
	\end{cases}\label{eq:Gvec}
\end{equation}
Given the definition $\bm{y}_{\bm{\delta} \mid \mathcal{M}}(t) = \bm{x}(t)$, Eq. \eqref{eq:Gvec} is for $a = 0$ and $\delta_i > 0$ equivalent to 
\begin{equation*}
	y_{0 \mid i}(t + 1) = G_i\left(\bm{x}(t)\right)  x_i(t).
\end{equation*}
For the case of no maturation delay with $\delta_i = 0$ such that $y_{0 \mid i}(t) = y_{\delta_i \mid i}(t) = x_i(t)$, Eq. \eqref{eq:Gvec} is equivalent to
\begin{equation*}
	y_{0 \mid i}(t + 1) = \sigma_{0 \mid i} x_i(t) +  G_i\left(\bm{x}(t)\right) x_i(t).
\end{equation*}
Eq. \eqref{eq:Gvec} then collapses to the equivalent delayed system,
\begin{equation}
	x_i(t + 1) = \sigma_{\delta_i \mid i} x_i(t) +  x_i(t - \delta_i)G_i(\bm{x}(t - \delta_i))\prod_{a = 0}^{\delta_i - 1} \sigma_{a \mid i}, \quad i \in \mathcal{M},\label{eq:collapsed}
\end{equation}
Eq. \eqref{eq:DDeq} is by Eq. \eqref{eq:FG} equivalent to Eq. \eqref{eq:collapsed}. Let $\bm{y}^\ast$ denote a positive equilibrium such that $\bm{y}^\ast = \bm{h}(\bm{y}^\ast) > \bm{0}$.  Note that the adult age class equilibrium abundances of Eq. \eqref{eq:Gvec} are equivalent to those of Eq. \eqref{eq:collapsed}, such that $\bm{y}^\ast_{\bm{\delta} \mid \mathcal{M}} = \bm{x}^\ast$, because $\bm{y}_{\bm{\delta} \mid \mathcal{M}}(t) = \bm{x}(t)$ by definition. 

The construction of Eq. \eqref{eq:Gvec} generalises the approach of \citet{Fisher1984} to the multispecies case, where adult reproduction through $G_i$ contributes to the recruits $y_{0 \mid i}(t + 1)$ of species $i$.  If the age classes are not of explicit interest then the assignment $\sigma_{a \mid i} = 1$ for $a \in \{0, \ldots, \delta_i - 1\}$ and $i \in \mathcal{M}$ in Eq. \eqref{eq:Gvec} is an obvious choice as used in the multispecies context by \citet{Fisher1984sys}.  The choice of how to specify the immature age classes is a matter of interpretation that may for example depend on whether immature individuals should be explicit and observable or implicit and latent.  Eq. (8) has been analysed for single species ($m = 1$) to establish local stability conditions by Clark (1976). For the single species case, \citet{Fisher1984} also construct a Lyapunov function on the scalar equation of Eq. \eqref{eq:collapsed} with  $m = 1$ to provide sufficient conditions of global stability for certain models.  Sufficient global stability conditions for the positive equilibrium \citep{Sacker2011} and boundary equilibria \citep{Jiang2017, Hou2021} of specific multispecies non-delayed versions of Eq. \eqref{eq:collapsed} with competitive density dependent growth functions have previously been presented. 

Here, sufficient local stability conditions are provided for the general multispecies system with delays. The Jacobian matrix of Eq. \eqref{eq:Gvec} evaluated at equilibria derived from Eq. \eqref{eq:collapsed} for the adult compartments is investigated in Section \ref{sec:local}. A sufficient condition for local asymptotic stability is then derived for the case of non-zero adult equilibrium abundances (Section \ref{sec:suff_gen}). The case of boundary equilibria is investigated in Section \ref{sec:bound_gen}.  A detailed examination is then provided for a multispecies competitive system of particular interest (Section \ref{sec:multi_comp}). 

\subsection{Local Asymptotic Stability}\label{sec:local}

Applying Eq. \eqref{eq:Clark} shows that for the single species case ($m = 1$), a sufficient condition for the local stability of Eq. \eqref{eq:collapsed} at an equilibrium $x_1^\ast$ is
\begin{equation}
	 \left| G_1(x_1^\ast) + x_1^\ast  \left. \frac{d G_1(x_1(t))}{dx_1(t)}\right|_{x_1(t) = x_1^\ast}  \right| \times \prod_{a = 0}^{\delta_1 - 1} \sigma_{a \mid i} < z_1.\label{eq:suff1}
\end{equation}
The above sufficient condition does not depend on the delay except through the product of the survival probabilities of the immature stages. Necessary and sufficient conditions for local stability, however, depend on the delay with closed-form expressions only available for certain choices of functional forms and parameter values \citep{Clark1976, Fisher1984}. It will be shown that a closed form expression similar to Eq. \eqref{eq:suff1} provides a sufficient condition for local stability in the multispecies delayed system of Eqs. \eqref{eq:Gvec} and \eqref{eq:collapsed}.

Let $\bm{J}$ denote the  matrix of partial derivatives of Eq. \eqref{eq:Gvec} with dimensions $(m + \sum_{i \in \mathcal{M}} \delta_i) \times (m + \sum_{i \in \mathcal{M}} \delta_i)$, such that the $(k, l)$ entry of $\bm{J}$ is $J_{k l} = \partial h_k\left(\bm{y}(t)\right) / \partial y_l(t)$. An equilibrium $\bm{y}^\ast$ of the multispecies age classes of the delayed system of Eq. \eqref{eq:Gvec} is locally asymptotically stable if the moduli of all eigenvalues of the Jacobian matrix  evaluated at the equilibrium, denoted by the matrix $\bm{J}^\ast$, are less than one. The spectral radius of a square matrix is defined as the maximum magnitude of all its eigenvalues. The equilibrium  $\bm{y}^\ast$ of Eq. \eqref{eq:Gvec} is therefore locally asymptotically stable if and only if the spectral radius of $\bm{J}^\ast$ is less than one, $\rho(\bm{J}^\ast) < 1$. 

For species $i$ with maturation delay $\delta_i > 0$, the partial derivatives of Eq. \eqref{eq:Gvec} evaluated at equilibrium are then
\begin{equation}
	\left.\frac{\partial h_{a \mid i}\left(\bm{y}(t)\right)}{\partial y_{b \mid j}(t)}\right|_{\bm{y}(t) = \bm{y}^\ast} =
	\begin{cases}
		\sigma_{a \mid i} & \textrm{if } i = j, \, a = b - 1, \, b = 1, \ldots, \delta_i,\\
		\sigma_{\delta_i \mid i} & \textrm{if } i = j,\, a = b = \delta_i\,,\\
		G_i(\bm{x}^\ast) + x_i^\ast  \left. \frac{\partial G_i(\bm{x}(t))}{\partial x_i(t)}\right|_{\bm{x}(t) = \bm{x}^\ast}   & \textrm{if } i = j,\, a = 0, \, b = \delta_j,\\
		\left. x_i^\ast \frac{\partial G_i(\bm{x}(t))}{\partial x_j(t)}\right|_{\bm{x}(t) = \bm{x}^\ast} &
		\textrm{if } i \neq j,\, a = 0, \, b = \delta_j,\\
		0 & \textrm{otherwise.}
	\end{cases}\label{eq:Gpdx}
\end{equation}
For species $i$ without delayed maturity ($\delta_i = 0$), the partial derivatives are
\begin{equation}
	\left.\frac{\partial h_{0 \mid i}\left(\bm{y}(t)\right)}{\partial y_{b \mid j}(t)}\right|_{\bm{y}(t) = \bm{y}^\ast} =
	\begin{cases}
		\sigma_{0 \mid i} + G_i(\bm{x}^\ast) + x_i^\ast  \left. \frac{\partial G_i(\bm{x}(t))}{\partial x_i(t)}\right|_{\bm{x}(t) = \bm{x}^\ast} & \textrm{if } i = j, \, b = \delta_j,\\
		\left. x_i^\ast \frac{\partial G_i(\bm{x}(t))}{\partial x_j(t)}\right|_{\bm{x}(t) = \bm{x}^\ast} & \textrm{if } i \neq j,\, b = \delta_j,\\
		0 & \textrm{otherwise}.
	\end{cases}	\label{eq:Gpd0}
\end{equation}
Evidently, only the zero age classes ($a = 0$) in Eq. \eqref{eq:Gpdx} depend on the equilibrium values.
The Jacobian matrix evaluated at the equilibrium $\bm{y}^\ast$ is 
\begin{equation}
	\bm{J}^\ast = 
	\begin{bmatrix}
		\bm{J}_{11}^\ast & \bm{J}_{12}^\ast & \cdots & \bm{J}_{1m}^\ast \\
		\bm{J}_{21}^\ast & \bm{J}_{22}^\ast & \cdots & \bm{J}_{2m}^\ast \\
		\vdots & \vdots & \ddots & \vdots \\
		\bm{J}_{m1}^\ast & \bm{J}_{m2}^\ast & \cdots & \bm{J}_{mm}^\ast
	\end{bmatrix}\label{eq:Jeq}
\end{equation}
that  consists of block matrices defined by
\begin{equation*}
	\bm{J}_{ij} = \left\{\begin{aligned}
	&&\begin{bmatrix}
		0 & 0 & \cdots & 0 & 0 & \sigma_{0 \mid i} + G_i(\bm{x}^\ast) + x_i^\ast  \left. \frac{\partial G_i(\bm{x}(t))}{\partial x_i(t)}\right|_{\bm{x}(t) = \bm{x}^\ast}  \\
		\sigma_{0 \mid i} & 0 & \cdots & 0 & 0 & 0 \\
		\vdots & \ddots & \ddots & & & \vdots \\
		0 & 0 & \cdots & \sigma_{\delta_i - 2 \mid i} & 0 & 0 \\
		0 & 0 & \cdots & 0 & \sigma_{\delta_i - 1 \mid i} & \sigma_{\delta_i \mid i}
	\end{bmatrix} & \quad \textrm{if } i = j, \\
	&&\begin{bmatrix}
		\bm{0} & \left. x_i^\ast \frac{\partial G_i(\bm{x}(t))}{\partial x_j(t)}\right|_{\bm{x}(t) = \bm{x}^\ast} \\
		\bm{0} & \bm{0}\\
	\end{bmatrix} \hspace{75pt} & \quad \textrm{if } i \neq j,
	\end{aligned}\right.
\end{equation*}
where the block matrix in the second case has all zero entries except for the last entry in its first row if $\delta_i > 0$ and, by Eqs. \eqref{eq:Gpdx} and \eqref{eq:Gpd0},
\begin{equation*}
	\bm{J}_{ij} = \left\{\begin{aligned}
		&&\begin{bmatrix}
			\sigma_{0 \mid i} + G_i(\bm{x}^\ast) + x_i^\ast  \left. \frac{\partial G_i(\bm{x}(t))}{\partial x_i(t)}\right|_{\bm{x}(t) = \bm{x}^\ast}
		\end{bmatrix} & \quad \textrm{if } i = j, \\
		&&\begin{bmatrix}
			\left. x_i^\ast \frac{\partial G_i(\bm{x}(t))}{\partial x_j(t)}\right|_{\bm{x}(t) = \bm{x}^\ast}
		  \end{bmatrix} \hspace{35pt} & \quad \textrm{if } i \neq j,
	\end{aligned}\right.
\end{equation*}
if $\delta_i = 0$ otherwise.

\subsection{Sufficient Condition for Local Stability of Coexistence Equilibrium}\label{sec:suff_gen}

The graph of Eq. \eqref{eq:Jeq} is used to construct a sufficient condition for the local asymptotic stability of the equilibrium. This result builds on a well-known result that bounds the location of the eigenvalues. Let $s(\bm{B})$ denote the spectrum of matrix $\bm{B}$, which is the set of eigenvalues of $\bm{B}$, that is the point set of values of $\bm{v}$ that satisfy $\det(\bm{B} - \bm{v}\bm{I}) = 0$.  The Ger{\u s}gorin theorem \citep[see, e.g.,][]{HJ1985} provides a geometric argument for bounding the spectrum of eigenvalues in the complex plane.  
\begin{theorem}[Ger{\u s}gorin]\label{thm:G}
	For $n \times n$ matrix $\bm{B}$, let
	\begin{equation*}
		R_i(\bm{B}) = \sum_{j \neq i} \left|B_{ij}\right|\,, \quad \forall i\,,
	\end{equation*}
	denote the deleted absolute row sums. The eigenvalues of $\bm{B}$ are located in the region formed by the union of $n$ discs such that
	\begin{equation*}
		s(\bm{B}) \subset \cup_{i} \left\{v: \left|v - B_{ii}\right| \leq R_i(\bm{B})\right\}.
	\end{equation*}
\end{theorem}
Ger{\u s}gorin's theorem  establishes that the eigenvalues of $\bm{B}$ are located within the Ger{\u s}gorin region. 
The discs that form a Ger{\u s}gorin region are called Ger{\u s}gorin discs and the corresponding boundaries of the discs are Ger{\u s}gorin circles.

Direct application of the Ger{\u s}gorin theorem to $\bm{J}^\ast$ does not necessarily bound the Ger{\u s}gorin discs within unit distance of the origin of the complex plane. The deleted absolute row sums that correspond to rows with the adult survival probabilities, $\sigma_{\delta_i \mid i}$ for $i \in \mathcal{M}$ in Eq. \eqref{eq:Jeq}, are not necessarily less than $\sigma_{\delta_i \mid i}$. If any of these sums exceed one then at least one Ger{\u s}gorin circle extends beyond unit distance of the origin, and the Ger{\u s}gorin theorem  then fails to bound the spectral radius to less than one. For example, this occurs in the case of a simple lag in adult emergence, where $\sigma_{a \mid i} = 1$ for $a = 0, \ldots, \delta_i - 1$. Application of the theorem to the transposition of the Jacobian matrix would give rise to consideration of the deleted absolute column sums. However, deleted absolute column sums of $\bm{J}^\ast$ that include the partial derivatives of the per capita growth function must similarly account for the adult stage daily survival probabilities.

More can be said, however, if the graph of $\bm{J}^\ast$ is also taken into account. The following terminology is needed \citep[][Ch. 6]{HJ1985}; see also \citet{Patten1985} for an ecological application. The directed graph of $n \times n$ matrix $\bm{J}^\ast$ has a directed arc (edge) between a pair $P_i$ and $P_j$ of the $n$ nodes $P_1, \ldots, P_n$ if and only if ${J}^\ast_{ij} \neq 0$. A directed path is a sequence of arcs $P_{i_1}P_{i_2}$, $P_{i_2}P_{i_3}$, $P_{i_3}P_{i_4}$, $\ldots$, with length defined as the number of arcs in the sequence. A cycle is a directed path that starts and ends at the same node such that no other node appears more than once; this is sometimes named a simple (directed) cycle.  A nontrivial cycle $\gamma$ excludes cycles of length one. If there exists a directed path of finite length that begins at $P_i$ and ends at $P_j$ for every pair $\{ P_i, P_j \}$ of distinct nodes in the directed graph of $\bm{J}^\ast$, then the directed graph is then said to be strongly connected.  

The definitions of (weakly or strongly) connected graphs and (weakly) irreducible matrices are also required. Let $|\bm{B}|$ denote the matrix with entries equal to the absolute value of the entries of $\bm{B}$. A matrix $\bm{B}$ with all positive (nonnegative) entries is denoted by $\bm{B} > \bm{0}$ ($\bm{B} \geq \bm{0}$). \citet[][Ch. 6]{HJ1985} show the equivalence between the following statements for an $n \times n$ matrix $\bm{B}$:
\begin{enumerate}
	\item $\bm{B}$ is irreducible,
	\item $(\bm{I} + |\bm{B}|)^{n - 1} > \bm{0}$, and
	\item the directed graph of $\bm{B}$ is strongly connected.
\end{enumerate}
Let $\alpha_1, \ldots, \alpha_r$ form a partition of $\{1, \ldots, n\}  = \cup_{i= 1}^r \alpha_i$ such that $\alpha_{i} \cap \alpha_j = \emptyset$ for $i \neq j$. Let the submatrices $\bm{B}(\alpha_i, \alpha_j)$ indexed by rows $\alpha_i$ and columns $\alpha_j$ of $\bm{B}$ with $1 \leq i, j \leq r$ form a partition of the matrix $\bm{B}$. The matrix $\bm{B}$ is reducible if either $n = 1$ and $\bm{B} = \bm{0}$, or a permutation $\bm{P}$ exists to triangular block form, 
\begin{equation}
	\bm{P}\bm{B}\bm{P}^{-1} = 
	\begin{bmatrix}
		\bm{B}({\alpha_1, \alpha_1}) & \bm{0} & \bm{0} & \bm{0} \\
		\bm{B}(\alpha_2, \alpha_1) & \bm{B}(\alpha_2, \alpha_2) & \bm{0} & \bm{0} \\
		\vdots & \vdots & \ddots & \vdots \\
		\bm{B}(\alpha_r, \alpha_1) & \bm{B}(\alpha_r, \alpha_2) & \cdots & \bm{B}(\alpha_r, \alpha_r)
	\end{bmatrix},\label{eq:reducible}
\end{equation}
where the diagonal blocks $\bm{B}(\alpha_w, \alpha_w)$ for $w \in \{1, \ldots, r\}$ are square, and either irreducible or a zero entry. These diagonal blocks correspond to the classes of $\bm{B}$, where a class is a set of nodes that communicate \citep{Berman1994}. Two nodes $P_i$ and $P_j$, $i \neq j$, communicate if there is a directed path from $P_i$ to $P_j$, and also from $P_j$ to $P_i$. An irreducible matrix $\bm{B}$ is a triangular block matrix with a single block, and an irreducible matrix is not reducible.

Additionally, a matrix $\bm{B}$ is said to be weakly irreducible if $\bm{C} = (\bm{I} + |\bm{B}|)^{n - 1}$ has at least one nonzero off-diagonal entry ${C}_{ij}$, $i \neq j$, in each row $i = 1, \ldots, n$ such that $C_{ji}$ is also nonzero. The matrix $\bm{B}$ is weakly irreducible if and only if its directed graph is weakly connected such that a nontrivial cycle exists between each node and another node. Note that an irreducible $n \times n$ matrix $\bm{B}$ with $n > 1$ is weakly irreducible. 

The following theorem \citep{Brualdi1982} uses the weakly connected property. Let $C(\bm{B})$ denote the set of all nontrivial cycles $\gamma$ in the directed graph of $\bm{B}$. 
\begin{theorem}[Brualdi]\label{thm:Brualdi}
	If the $n\times n$ matrix $\bm{B}$ is weakly irreducible, then 
	\begin{equation*}
		s(\bm{B}) \subset \underset{\gamma \in C(\bm{B})}{\bigcup} \left\{v: \prod_{P_i \in \gamma } |v - B_{ii}| \leq \prod_{P_i \in \gamma} R_i(\bm{B}) \right\}.
	\end{equation*}
\end{theorem}
If a cycle $\gamma$ of length $k > 1$ includes nodes $P_{i_1}P_{i_2}$, $\ldots$, $P_{i_k}P_{i_{k + 1}}$ with $P_{i_1} = P_{i_{k + 1}}$ then the index $i = i_1, i_2, \ldots, i_k$ in the products of the above theorem.
\begin{theorem}\label{prop:suff}
	 Assume that the system of Eq. \eqref{eq:Gvec} with adult stages $\bm{x}(t) = \bm{y}_{\bm{\delta} \mid \mathcal{M}}(t)$ has an interior equilibrium, $\bm{x}^\ast = \bm{y}^\ast_{\bm{\delta} \mid \mathcal{M}} > \bm{0}$. Let the Jacobian matrix of Eq. \eqref{eq:Jeq} with dimensions  $(m + \sum_{i \in \mathcal{M}} \delta_i) \times (m + \sum_{i \in \mathcal{M}} \delta_i)$ be expressed in the triangular block form of Eq. \eqref{eq:reducible} with partition $\alpha_1, \ldots, \alpha_r$. Let $\mathcal{D}_w \subseteq \mathcal{M}$ denote the set of species with nodes in $\alpha_w$. If, for all $i \in \mathcal{D}_w$ and $w \in \{1, \ldots, r\}$,
	\begin{equation*}
		\left\{\left|\left( G_i(\bm{x}^\ast) + x_i^\ast  \left. \frac{\partial G_i(\bm{x}(t))}{\partial x_i(t)}\right|_{\bm{x}(t) = \bm{x}^\ast} \right) \right|  + \sum_{j \in \mathcal{D}_w\backslash i} \left|\left(\left. x_i^\ast\frac{\partial G_i(\bm{x}(t))}{\partial x_j(t)}\right|_{\bm{x}(t) = \bm{x}^\ast}\right) \right| \right\} \prod_{a = 0}^{\delta_i - 1} \sigma_{a \mid i} < z_i, 
	\end{equation*}
	then the equilibrium is locally asymptotically stable.
\end{theorem}
The proof of the above theorem is given in Appendix \ref{app:suff}.

Theorem \ref{prop:suff} is the multispecies generalisation of Eq. \eqref{eq:suff1}. The proposition does not explicitly depend on the lags $\bm{\delta}$ except through the product of survival probabilities of the immature stages. 
Theorem \ref{prop:suff} uses Theorem \ref{thm:Brualdi} that depends on the weakly irreducible condition. 
The weakly irreducible condition for $\bm{J}^\ast$ holds by Eq. \eqref{eq:Jeq} if $\delta_i \geq 1$ for all $i \in \mathcal{M}$. 
An irreducible $\bm{J}^\ast$ is also weakly irreducible. 
For completeness, a sufficient condition for irreducibility and the strongly connected property follows.
\begin{lemma}\label{lemma:Girr}
	If the partial derivatives of the zero age classes with the mature age classes in Eqs. \eqref{eq:Gpdx} and \eqref{eq:Gpd0} evaluated at equilibrium are all non-zero, then $\bm{J}^\ast$ is irreducible.
\end{lemma}
The proof is in Appendix \ref{app:Girr}.

\subsection{Sufficient Conditions for Local Stability of Boundary Equilibria}\label{sec:bound_gen}

Let $\bm{\hat{x}} = [\bm{x}^\circ, \bm{x}^\bullet]^\top$ denote a boundary equilibrium, where $\bm{x}^\circ = \bm{0}$ is an $r$-dimensional vector with $r \in \{1, 2, \ldots, m\}$. The boundary equilibria are those where the positive equilibrium abundances in $\bm{x}^\bullet > \bm{0}$ form a $p$-dimensional vector, $p = m - r$. If $\bm{x}^\circ$ is $m$-dimensional then $\bm{\hat{x}} = \bm{0}$ is the trivial equilibrium. A sufficient condition for the local asymptotic stability is provided by application of Theorem \ref{prop:suff} to the Jacobian matrix evaluated at $\bm{\hat{x}}$. However, if the per capita population growth functions satisfy a certain condition then more can be said.

Let $\bm{\hat{J}}$ denote the Jacobian matrix of Eq. \eqref{eq:Jeq} evaluated at an equilibrium  where at least one of the species has zero abundance. Let $\bm{{J}^\circ}$ and $\bm{{J}^\bullet}$ denote the block submatrices on the main diagonal of $\bm{\hat{J}}$  conformally partitioned with  $\bm{\hat{x}}$ such that the first $r$ rows correspond to the $r$ species with zero equilibrium abundances.
\begin{theorem}\label{prop:suff_alt}
	 Assume that Eq. \eqref{eq:Gvec} for $i \in \mathcal{M}$ has the property that
	 \begin{equation*}
	 	\hat{x}_i = x_i^\circ = 0 \implies \hat{x}_i G_i(\bm{\hat{x}}) = 0, 
	 \end{equation*}
	 and assume $\rho(\bm{{J}^\bullet}) < 1$. The equilibrium $\bm{\hat{x}}$ is locally asymptotically stable if 
	 \begin{equation*}
	 	\left| G_i(\bm{\hat{x}}) \right|  \prod_{a = 0}^{\delta_i - 1} \sigma_{a \mid i}  < z_i, \quad \forall i \in \{k: \hat{x}_k = x_k^\circ = 0\}.
	 \end{equation*}
\end{theorem}
The proof of the above theorem is found in Appendix \ref{app:suff_alt}.

The condition on the per capita growth function $G_i$ of  an introduced or invading species with zero equilibrium abundance $x_i^\circ$  says that its population cannot grow in the absence of any reproducing individuals for  species $i$. The assumption on the spectral radius $\rho(\bm{{J}^\bullet})$ corresponds to a locally stable interior equilibrium $\bm{x}^\bullet$ for the $p$ species with non-zero equilibrium abundances in the absence of the $r$ species with zero equilibrium abundances, $\bm{x}^\circ$.

\section{Multispecies Competition with Delayed Maturation}\label{sec:multi_comp}

For an important specific example of Eqs. \eqref{eq:Gvec} and \eqref{eq:collapsed} with $\bm{y}_{\bm{\delta} \mid \mathcal{M}}(t) = \bm{x}(t)$, consider the per capita density dependent growth function defined by
\begin{equation}
	G_i\left(\bm{x}(t - \delta_i)\right) = \frac{{\lambda}_i }{1 + q_i^{-1} \sum_{k \in \mathcal{M}} A_{ik} x_k(t - \delta_i)}, \quad i \in \mathcal{M},\label{eq:G}
\end{equation}
where $\bm{A} \geq \bm{0}$ is an $m \times m$ matrix with positive main diagonal entries and nonnegative off-diagonal entries. The matrix $\bm{A}$ defines the intraspecific and interspecific competitive interactions that intensify with greater numbers of reproducing adults both within and among species. By construction, a positive entry $A_{ij} > 0$ indicates a negative effect of species $j$ on species $i$.  Each entry  of the nonnegative matrix $\bm{A}$ defines the strength  of competitive density dependence exerted by species $j$  after duration $\delta_i$ in the immature life history stage of species $i$.  The off-diagonal entry $A_{ij}$ defines the strength of density dependence of species $j$ on species $i$ relative to the intraspecific density dependence, $A_{ii}$, on the  species $i$. The $m$-dimensional vector $\bm{\lambda} = [\lambda_1, \ldots, \lambda_m]^\top$ of density independent fecundity rates specifies the per capita number of new recruits generated by the adults of each species in the absence of density dependence that would occur if $\bm{A} = \bm{0}$. The $i$\textsuperscript{th} row of $\bm{A}$ assesses the relative magnitudes of each species on the population growth of species $i$. The parameter $q_i > 0$ determines the overall impact of both intraspecific and interspecific density dependence on the population growth rate of species $i$.

\begin{remark}\label{remark:q}
	Without loss of generality, the specification $A_{ii} = 1$ for $i \in \mathcal{M}$ in the CDM model defines $q_i^{-1}$ as the strength of intraspecific density dependence in the denominator of Eq. \eqref{eq:G}. 
\end{remark}

The general delayed model with structure defined by Eqs. \eqref{eq:Gvec}, \eqref{eq:collapsed} and \eqref{eq:G} and $m \geq 1$ is for convenience referred to here as the Competitive Delayed Multispecies (CDM) model. The nondelayed version of Eq. \eqref{eq:collapsed} with $\bm{\sigma} = \bm{0}$ and per capita growth function defined by Eq. \eqref{eq:G}  is a generalisation of models that have a long history in mathematical biology. For $m = 1$, it is the influential Beverton-Holt model \citep{BH1957}. For $m = 2$, it is the Leslie-Gower model \citep{Leslie1958}. For $m > 1$, the non-delayed model is known as the competitive multispecies Leslie-Gower model \citep{Sacker2011, Jiang2017}. A single species delayed version appears in \citet{Fisher1984} and a multispecies competitive system with common delay $\delta_i = \delta$ for all $i \in \mathcal{M}$ appears in \citet{Hosack2023}.

Before proceeding with multispecies analysis, the stability properties for the case $m = 1$ are now provided. If $\lambda_1 > 1$ in the single species model ($m = 1$) without delay ($\delta_1 = 0$), then the model is of  Beverton-Holt  form with a positive equilibrium, $ x_1^\ast = q_1 (\lambda_1 - 1)  / A_{11}$, that is globally asymptotically stable \citep{Elaydi2005}; if $0 < \lambda_1 < 1$ then the zero solution is asymptotically stable. The delayed version of this model with $\delta_1 > 0$ from Eqs. \eqref{eq:FG} and Eq. \eqref{eq:G} multiplies the fecundity rate by the product of juvenile stage survival probabilities. Application of the results of \citet{Fisher1984sys} for the case of maturation delay ($\delta_1 > 0$) also multiplies the fecundity rate  by a factor $z_1^{-1}$ in the corresponding non-delayed system of Eq. \eqref{eq:nondelayed_system} given the per capita growth function of Eq. \eqref{eq:G}.  Therefore, if $\lambda_1 \prod_{a = 1}^{\delta_1 - 1} \sigma_{a \mid 1} > z_1 $ in the delayed Beverton-Holt system, then its  equilibrium $x_1^\ast = q_1 (z_1^{-1}\lambda_1 \prod_{a = 1}^{\delta_1 - 1} \sigma_{a \mid 1} - 1)  / A_{11}$ is both positive and globally asymptotically stable. The stability conditions for the single species model with or without delayed maturation do not depend  on the parameters $q_1$ and $A_{11}$ that scale its positive equilibrium. However, unlike the non-delayed Beverton-Holt model, the stability condition for $\delta_1 > 0$ depends on both juvenile and adult stage survival probabilities.

\subsection{Coexistence Equilibrium}\label{sec:coexist}

The positive equilibrium  with species-specific delays is derived. Let the $m$-dimensional vector $\bm{x}(t)$ denote the  reproducing adult abundances for the set of species $\mathcal{M}$ at time $t$, such that $x_i(t) = y_i(t, \delta_i)$ for $i \in \mathcal{M}$. Let $\bm{x}^\ast > \bm{0}$ denote the interior equilibrium abundances of the reproducing adult age classes, and define the density independent rate of adult recruitment of species $i$ by $\tilde{\lambda}_i = \lambda_i\prod_{a = 0}^{\delta_i - 1} \sigma_{a \mid i}$. If $\delta_i = 0$ then $\tilde{\lambda}_i = \lambda_i$ (Section \ref{sec:model}). If the positive equilibrium exists, then Eqs. \eqref{eq:collapsed} and \eqref{eq:G} obtain 
\begin{equation}
	x_i^\ast = \sigma_{\delta_i \mid i} x_i^\ast +  \frac{\tilde{\lambda}_i x_i^\ast}{1 + q_i^{-1} \sum_{k \in \mathcal{M}} A_{ik} x_k^\ast}, \quad i \in \mathcal{M}.\label{eq:process}
\end{equation}
Using the daily probability of mortality $z_i = 1 - \sigma_{\delta_i \mid i}$ and rearranging terms  gives
$\sum_{j \in \mathcal{M}} A_{ij} x_j^\ast = \left(\frac{\tilde{\lambda}_i}{z_i} - 1\right)q_i$ for all $i \in \mathcal{M}$,
or equivalently
$\bm{A}\bm{x}^\ast = \textrm{diag}[\bm{\tilde{\lambda}} - \bm{z}]\textrm{diag}[\bm{z}]^{-1}\bm{q}$,
where $\textrm{diag}[\bm{v}]$ is the $d \times d$ diagonal matrix with main diagonal entries corresponding to the entries of a $d$-dimensional vector $\bm{v}$ and zeroes in the off-diagonal entries. For nonsingular $\bm{A}$, the equilibrium abundances are \begin{equation}
	\bm{x}^\ast = \bm{A}^{-1}\textrm{diag}\left[\bm{\tilde{\lambda}} - \bm{z}\right]\textrm{diag}[\bm{z}]^{-1}\bm{q}.\label{eq:xeq}
\end{equation}
A feasible interior equilibrium of Eq. \eqref{eq:process} for an arbitrary number of species $m \geq 1$ corresponds to a positive equilibrium $\bm{x}^\ast > \bm{0}$ that is unique by  Eq. \eqref{eq:xeq}. 

Often the density independent parameters $\bm{\tilde{\lambda}}$ and $\bm{z}$ that relate to life history parameters within a species have comparatively more information available than the more uncertain density dependent parameters $\bm{A}$ and $\bm{q}$ that govern relationships among species. What can be generally said about the feasibility of the coexistence equilibrium if the exact values of $\bm{q}$ are unknown or uncertain? The following proposition sets sufficient conditions on $\bm{A}$ such that the interior equilibrium is feasible if $\bm{q} > \bm{0}$ are free parameters.
\begin{lemma}\label{lemma:eq}
	If the feasible interior equilibrium exists, $\bm{x}^\ast > \bm{0}$, given $\bm{\tilde{\lambda}} > \bm{z} > \bm{0}$ and nonsingular $\bm{A}$ with all positive entries on the main diagonal, then $\bm{q} > \bm{0}$ exists.
\end{lemma}
The proof is found in Appendix \ref{app:eq}.

The condition $\bm{\tilde{\lambda}} > \bm{z}$ corresponds to positive population growth rates in the absence of competition, and the condition $\bm{z} > \bm{0}$ corresponds to non-negligible density independent mortality of mature adults. The condition of all positive main diagonal entries of $\bm{A}$ corresponds to the presence of intraspecific competition for all species. Lemma \ref{lemma:eq} then guarantees that a positive $\bm{q}$ exists for any feasible interior equilibrium $\bm{x}^\ast > \bm{0}$. 

\subsection{Jacobian Matrix at Coexistence Equilibrium}\label{sec:Jco}

Sufficient conditions for the global stability of the positive equilibrium of the non-delayed version of the CDM model have been presented \citep{Sacker2011, Jiang2017, Hou2021}. For the single species case $m = 1$ with delay $\delta_1 > 0$, sufficient conditions for the global stability of the positive equilibrium $x_1^\ast = q_1(\lambda_1 - z_1)/z_1$ have been provided \citep{Fisher1984}. The positive equilibrium $\bm{x}^\ast > 0$ of the general CDM model, if it exists, is locally asymptotically stable if the moduli of all eigenvalues of the Jacobian matrix of Eq. \eqref{eq:Jeq} evaluated at the feasible interior equilibrium are less than one. For species $i$ with maturation delay $\delta_i > 0$,  the partial derivatives are
\begin{equation}
	\frac{\partial h_{a \mid i}\left(\bm{y}(t)\right)}{\partial y_{b \mid j}(t)} =
	\begin{cases}
		\sigma_{a \mid i} & \textrm{if } i = j, \, a = b - 1, \, b = 1, \ldots, \delta_i,\\
		\sigma_{\delta_i \mid i} & \textrm{if } i = j,\, a = b = \delta_i,\vspace{3mm}\\
			\dfrac{{{\lambda}}_i\left(1 - \dfrac{q_i^{-1}A_{ii} y_{\delta_i \mid i}(t)}{1 + q_i^{-1}\sum_{k \in \mathcal{M}} A_{ik} y_{\delta_k \mid k}(t)}\right)}{1 + q_i^{-1} \sum_{k \in \mathcal{M}} A_{ik} y_{\delta_k \mid k}(t)} & \textrm{if } i = j,\, a = 0, \, b = \delta_j,		\vspace{3mm}\\
		\dfrac{-{{\lambda}}_i q_i^{-1} A_{ij} y_{\delta_i \mid i}(t)}{\left(1 + q_i^{-1} \sum_{k \in \mathcal{M}} A_{ik} y_{\delta_k \mid k}(t)\right)^2} &
		\textrm{if } i \neq j,\, a = 0, \, b = \delta_j,		\vspace{3mm}\\	
		0 & \textrm{otherwise.}
	\end{cases}\label{eq:pdx}
\end{equation}
For species $i$ without delayed maturity ($\delta_i = 0$), the partial derivatives are
\begin{equation}
	\frac{\partial h_{0 \mid i}\left(\bm{y}(t)\right)}{\partial y_{a \mid j}(t)} =
	\begin{cases}
		\sigma_{0 \mid i} + \dfrac{{{\lambda}}_i\left(1 - \dfrac{q_i^{-1}A_{ii} y_{0 \mid i}(t)}{1 + q_i^{-1}\sum_{k \in \mathcal{M}} A_{ik} y_{\delta_k \mid k}(t)}\right)}{1 + q_i^{-1} \sum_{k \in \mathcal{M}} A_{ik} y_{\delta_k \mid k}(t)} & \textrm{if } i = j,\, a = \delta_j\,,\vspace{3mm}\\
		\dfrac{-{{\lambda}}_i q_i^{-1} A_{ij} y_{0 \mid i}(t)}{\left(1 + q_i^{-1} \sum_{k \in \mathcal{M}} A_{ik} y_{\delta_k \mid k}(t) \right)^2} &
		\textrm{if } i \neq j,\, a = \delta_j,\vspace{3mm}\\
		0 & \textrm{otherwise.}
	\end{cases}\label{eq:pd0}
\end{equation}

Let $\bm{J}$ denote the Jacobian matrix of partial derivatives defined by Eqs. \eqref{eq:pdx} and \eqref{eq:pd0}. Note that Eq. \eqref{eq:xeq} is equivalent to
\begin{equation}
	1 + q_i^{-1} \sum_{j \in \mathcal{M}} A_{ij} x_j^\ast = \frac{\tilde{\lambda}_i}{z_i}, \quad i \in \mathcal{M}.
	\label{eq:invq}
\end{equation}
Assume that $\bm{A}$ is nonsingular and that intraspecific density dependence occurs for all species. Then a positive $\bm{q} > \bm{0}$ exists for a corresponding feasible interior equilibrium $\bm{x}^\ast$ by Lemma \ref{lemma:eq}. The parameters in $\bm{q}$ appear only in the zero age classes of Eqs. \eqref{eq:pdx} and \eqref{eq:pd0}. If $\bm{x}^\ast > \bm{0}$ in Eq. \eqref{eq:xeq} then Eq. \eqref{eq:invq} is applied so that the parameters $\bm{q}$ are implicit in the entries expressed as quotients of the equilibrium values. For the zero ($a = 0$) age class of each species $i \in \mathcal{M}$, Eqs. \eqref{eq:pdx}, \eqref{eq:pd0} and \eqref{eq:invq} yield partial derivatives evaluated at the interior equilibrium $\bm{x}^\ast$ of
\begin{equation}
	\left.\frac{\partial h_{0 \mid i}\left(\bm{y}(t)\right)}{\partial y_{\delta_j \mid j}(t)}\right|_{\bm{y}(t) = \bm{y}^\ast} =
	\begin{cases}
		z_i\left(\dfrac{\lambda_i}{\tilde{\lambda}_i} - \dfrac{\lambda_i(\tilde{\lambda}_i - z_i)}{\tilde{\lambda}_i^2}\dfrac{A_{ii} x_i^\ast}{\sum_{k\in\mathcal{M}} A_{ik}x_k^\ast}\right) 
		& \textrm{if } i = j  \textrm{ and } \delta_i > 0,\vspace{3mm}\\
		\sigma_{0 \mid i} + z_i\left(1 - \dfrac{({\lambda}_i - z_i)}{{\lambda}_i}\dfrac{A_{ii} x_i^\ast}{\sum_{k\in\mathcal{M}}	 A_{ik}x_k^\ast}\right) 
		& \textrm{if } i = j  \textrm{ and } \delta_i = 0,\vspace{3mm}\\
		-z_i \dfrac{\lambda_i(\tilde{\lambda}_i - z_i)}{\tilde{\lambda}_i^2} \dfrac{A_{ij} x_i^\ast}{ \sum_{k \in \mathcal{M}} A_{ik} x^\ast_k}  &
		\textrm{if } i \neq j,\vspace{3mm}\\
		0 & \textrm{otherwise.}
	\end{cases}\label{eq:pG}
\end{equation}
The Jacobian matrix $\bm{J}^\ast$ evaluated at the feasible interior equilibrium  is then given by Eq. \eqref{eq:Jeq}.

\subsection{Numerical Evaluation of Stability for All Coexistence Equilibria}\label{sec:all}

The derived $\bm{J}^\ast$ enables two useful applications without knowledge of $\bm{q}$ that are usefully applied in the empirical application of Section \ref{sec:estimation}: i)  the stability of an interior equilibrium can be calculated given $\bm{x}^\ast$, and ii) the stability of all possible feasible interior equilibria in fact only require consideration of the equilibrium relative abundances of the $m$ species on the $m - 1$ simplex. These results are useful because otherwise the typically unknown or uncertain $\bm{q}$ scales the magnitudes of the species abundances at a feasible interior equilibrium such that  $\bm{x}^\ast \in \mathbb{R}_+^m$. For i), the Jacobian matrix evaluated at the feasible interior equilibrium by Eq. \eqref{eq:pG} shows that the typically unknown  vector $\bm{q}$ is accounted for in the stability analysis given a positive equilibrium $\bm{x}^\ast$. For ii), the feasible equilibrium abundances appear as ratios in Eq. \eqref{eq:pG}. To explore the entire feasible space of the Jacobian evaluated at any interior equilibrium $\bm{x}^\ast > \bm{0}$, define the total abundance of an interior equilibrium by $x_{T}^\ast = \sum_k x_k^\ast$. Thus the relative abundance of the  species $k$ is $p_k = x_k^\ast/x_{T}^\ast > 0$, that is, $x_k^\ast = p_k x_{T}^\ast$ with $\sum_k p_k = 1$, such that ${A_{ij}x_j^\ast}/{\sum_{k \in \mathcal{M}} A_{ik} x_k^\ast} = {A_{ij} p_j}/{\sum_{k \in \mathcal{M}} A_{ik} p_k}$ for all $i, j \in \mathcal{M}$ in Eq. \eqref{eq:pG}. Numerical evaluation of the eigenvalues of $\bm{J}^\ast$ for all possible interior equilibria is then implemented through exploration of the simplex $\{p_i: i \in \mathcal{M}, \, p_i > 0, \, \sum_{k \in \mathcal{M}} p_k = 1\}$ as the feasible region given only parameters $\bm{A}$, $\bm{{\lambda}}$ and $\bm{\sigma}$. 

Note that there are infinitely many equilibria represented within the simplex.  Each point in the simplex corresponds to the equilibrium relative abundances of the species, or fractional contribution of each species to the total abundance aggregated over all species at the equilibrium. The total abundance of an interior equilibrium ${x}_T^\ast$ is a strictly positive real number. Hence the possible equilibria that correspond to a single point in the simplex form an infinite set, and this is true of every point in the simplex that together form an infinite set. However, the set of eigenvalues or spectrum $s(\bm{J}^\ast)$ for a given interior equilibrium $\bm{x}^\ast$  depends only on the species equilibrium relative abundances and not the total abundance. Therefore, the condition that the moduli of all eigenvalues of $\bm{J}^\ast$ are less than one can  be numerically checked to assess local asymptotic stability of any $\bm{x}^\ast > \bm{0}$ by consideration of its corresponding unique point in the simplex. 

\subsection{Sufficient Condition for Stability of Coexistence Equilibrium}\label{sec:suff}

\begin{proposition}\label{prop:CDM}
	Assume that $\bm{\tilde{\lambda}} > \bm{z} > \bm{0}$. If
	\begin{equation*}
		A_{ii} > \sum_{j\in\mathcal{M}\backslash i} A_{ij}, \quad \forall i \in \mathcal{M},
	\end{equation*}
	then any interior equilibrium $\bm{x}^\ast > \bm{0}$ of the CDM model is feasible with an associated $\bm{q} > \bm{0}$, and the interior equilibrium is locally asymptotically stable.
\end{proposition}
The proof of the proposition is in Appendix \ref{app:CDM}.

The constraint of Proposition \ref{prop:CDM} shows that if interspecific competition is not greater than intraspecific competition, then the coexistence equilibrium exists and is locally stable. The condition does not depend on the species-specific delays.

\subsection{Sufficient Condition for Stability of Alternative Equilibria}\label{sec:alt}

By the uniqueness of the interior equilibrium of Eq. \eqref{eq:xeq}, if it exists, the only other possible equilibria for the CDM model are the boundary equilibria and the trivial equilibrium.
\begin{corollary}\label{cor:CDMalt}
	Assume that the condition of Proposition \ref{prop:CDM} applies to guarantee stability of the interior portion $\bm{x}^\bullet > \bm{0}$, of the equilibrium, if present, such that $\rho(\bm{{J}^\bullet}) < 1$. Then a boundary equilibrium $\bm{\hat{x}} = [\bm{x}^\circ, \bm{x}^\bullet]^\top$, where $\bm{x}^\circ = \bm{0}$, or trivial equilibrium, where $\bm{\hat{x}} = \bm{x}^\circ = \bm{0}$, is locally asymptotically stable if
	\begin{equation*}
		\frac{\tilde{\lambda}_i - z_i}{z_i} < q_i^{-1} \sum_{k \in \mathcal{M}} A_{ik} \hat{x}_k = q_i^{-1} \sum_{k \in \{l: \hat{x}_l = x_l^\bullet \}} A_{ik} x_k^\bullet, \quad \forall i \in \{j: \hat{x}_j = x_j^\circ = 0\},
	\end{equation*}
	where the empty sum  in the case of the trivial equilibrium  is zero.
\end{corollary}
The proof is in Appendix \ref{app:CDMalt}.

The sufficient condition for the local stability of the interior equilibrium only depends on the equilibrium through the relative abundances of the species (Section \ref{sec:all}); however, the corresponding sufficient condition for the stability of the boundary equilibrium in the above corollary depends on the actual non-zero equilibrium abundances $\bm{x}^\bullet$ of the established species. The boundary or trivial equilibrium is locally asymptotically stable if the fecundity rate at the conclusion of the immature life history phase is less than the probability of adult mortality. 

\section{Applications}\label{sec:estimation}

The above properties of the CDM model are now used to provide estimates for unknown density dependent parameters $\bm{q}$ and $\bm{A}$. In Section \ref{sec:Pianka}, a popular descriptive approach to estimate niche overlap is summarised. If coerced to define the entries of $\bm{A}$, this descriptive approach endows a special structure such  that Lemma \ref{lemma:eq} applies, and so $\bm{q} > \bm{0}$ is then also available. However, the niche overlap method is not strongly linked to a dynamic model framework. Therefore, Section \ref{sec:optim} explores an alternative approach that estimates $\bm{A}$ and $\bm{q}$ based on the dynamic properties of the CDM model. The optimisation uses the spectral radius to determine the resilience of the ecosystem in terms of the relative dampening or amplification of perturbations away from the interior equilibrium. The level of resilience, which is maximised by minimising the spectral radius, is balanced against a maximisation of the average interspecific competition strength among species, which is destabilising by Proposition \ref{prop:CDM}. The idea is that i) observed species abundances are more likely to persist in ecosystems with greater resilience, and ii) species seek to increase competitive advantage over shared resources. The approach requires only the relative abundances of the species rather than actual abundances that can be more difficult to obtain \citep{Sinka2016}. For example, models of habitat overlap and segregation among species sometimes require genetic methods to identify species from subsamples of field populations \citep{TeneFossog2015}. Such analyses often use relative abundance because the actual abundances of the species are unobserved. The descriptive and objective approaches are compared for an example  in Section \ref{sec:app}.

\subsection{Relationship between CDM Model and Niche Overlap Indices}\label{sec:Pianka}

It has long been recognised that interspecific density dependence can be difficult to estimate. \citet[][p. 169]{BH1957} for example suggest that ``a true measure of competition requires considerable knowledge of the dynamics of the system''. The Leslie-Gower model was an early attempt to provide a mechanistic model with estimable density dependence for a two-species extension of the Beverton-Holt model of density dependent population growth \citep{Leslie1958}. Later, various measures of niche overlap indices were introduced into the ecological literature that were loosely based on the continuous Lotka-Volterra model \citep{May1975}. \citet{Pianka1974}, for example, proposes 
$\alpha_{ij} = \sum_{r \in \mathcal{R}} p_{ir}p_{jr} \left(\sum_{r \in \mathcal{R}}p_{ir}^2 \sum_{r \in \mathcal{R}}p_{jr}^2 \right)^{-1/2}$
as a measure of niche overlap between species $i$ and $j$, where $p_{kr}$ is the proportion of the resource $r$ from the set of resources $\mathcal{R}$ used by species $k$. The resulting $\alpha_{ij}$ are proposed as the ``canonical'' estimator of interspecific competitive coefficients \citep{May1975}, where in  a competitive community of $m$ species, it is assumed that $A_{ij} = \alpha_{ij}$ is a density dependent coefficient. The resulting $m \times m$ interspecific competition matrix $\bm{A}$ constructed by this method is symmetric and positive semidefinite \citep{May1975}. 

The CDM model benefits in two ways from assuming $\bm{\alpha} = \bm{A}$. First, the main diagonal entries of $\bm{A}$ equal one. The unknown $q_i^{-1}$ in the density dependent growth functions $G_i(\bm{x}(t - \delta_i))$ of Eq. \eqref{eq:G} are therefore interpreted as setting intraspecific density dependence on the per capita scale for species $i$ as in Remark \ref{remark:q}. Second, the rank of a symmetric matrix corresponds to the number of its non-zero eigenvalues. Therefore, if the competitive relationships or niches among the species differ such that the rows (columns) of $\bm{A}$ are linearly independent, then the matrix $\bm{A}$ has full rank $m$ and all positive eigenvalues. It would be reasonable for different species to interact with their environments differentially, and linear independence of rows (columns) of $\bm{A}$ would typically be expected. The matrix $\bm{A}$ is then nonsingular with intraspecific density dependence present for all species. The unique interior equilibrium then exists by Lemma \ref{lemma:eq} if also $\bm{\tilde{\lambda}} > \bm{z}$. So the choice of parameterisation of $\bm{A}$ by the Pianka index guarantees a feasible interior equilibrium for the CDM model.

Yet \citet{Pianka1974} explicitly cautions that $\bm{\alpha}$ only measures ``niche overlap'', not competition, because species can adopt strategies to decrease competitive interactions even as their niche overlap increases. Another criticism of assuming $\bm{\alpha} = \bm{A}$ is that the coefficients $\bm{\alpha}$ are typically based on observations of relative abundances among species, whereas the unobserved competitive coefficients $\bm{A}$ are the real measure of interspecific and intraspecific density dependence on population growth rates. Thus, a field survey that shows strong positive correlation of relative abundances among species across sites may suggest strong competition because $\bm{\alpha}$ has large off-diagonal entries, whereas in fact the species may exploit distinct non-overlapping microhabitats within each site. In this example, the competition matrix $\bm{A}$ is a diagonal matrix that reflects the absence of interspecific competition so that clearly $\bm{\alpha} \neq \bm{A}$.

\subsection{Optimised Competition}\label{sec:optim}

The results available for the CDM model suggest an alternative approach to estimating the strength of competition.
Estimates for density independent adult mortality rates $\bm{z}$ and per capita reproduction rates $\bm{\tilde{\lambda}}$, as defined in the delayed maturity model, are generally available in the literature, but estimates of the density dependent parameters $\bm{q}$ and $\bm{A}$ are comparatively lacking. Estimates of the former parameters often depend on a mix of field observations or laboratory studies, and so are more attainable than the latter parameters that determine the magnitude of density dependence. Assume that an estimate of the adult abundances $\bm{\bar{x}} > \bm{0}$ is available for the interior equilibrium, so that  $\bm{\bar{x}} \approx \bm{x}^\ast$. The estimate $\bm{\bar{x}}$ may for example be derived from estimated mean adult abundances of the species. Here, an estimate of the competition matrix $\bm{A}$ is derived based on the stability properties of the dynamic system. This ecologically and dynamically motivated estimate is available for the CDM model in the absence of estimates for $\bm{q}$. The optimisation approach can also be viewed as a hypothesis for how competing species might set competitive interaction strengths at a stable, and potentially observable, configuration.

Whereas the condition for local asymptotic stability depends on the comparison of the spectral radius with the threshold of one, the following interpretation also considers the magnitude. This conceptual approach uses the spectral radius to assess the amplification potential for perturbations away from equilibrium, and is conceptually based on the analogous construction for the reproduction number of nonnegative populations \citep{Diekmann2000}. The magnitude of the spectral radius is also used in ecology as a measure of the ``resilience'' of a stable equilibrium in a system of difference equations \citep{Beddington1976}.  Resilience is a measure of species persistence in ecosystems subject to change and disturbance \citep{Holling1973} and  is defined as the rate of return to a stable equilibrium or the inverse of the return time after a perturbation \citep{Nakajima1989}. \citet{Beddington1976} defines the asymptotic return time to a stable equilibrium as
\begin{equation*}
	T_R = \left[1 - \rho\left(\bm{J}^\ast\right)\right]^{-1}.
\end{equation*}
A system of species with greater resilience and shorter return time $T_R$ is more likely to persist in the face of perturbations with corresponding decreased risk of species extinctions \citep{Beddington1976}. 

Let $\bm{\phi}$ denote the vector of deviations for all age classes of all species from the interior equilibrium caused by a local perturbation. The magnitude of this deviation can be summarised by the vector norm $||\bm{\phi}||_1 = \sum_{i} |\phi_i| $. The corresponding matrix norm for a $n \times n$ square matrix $\bm{W}$ is $|| \bm{W} ||_1 = \sum_{i, j = 1}^n |W_{ij}| = \max_{\bm{\phi} \neq \bm{0}} || \bm{W} \bm{\phi} ||_1 / || \bm{\phi} ||_1$ that  assesses the maximum amplification of the deviation from the interior equilibrium. The spectral radius of $\bm{W}$ for a matrix norm $||\cdot||$ is defined by $\rho(\bm{W}) = \lim_{t \rightarrow \infty} || \bm{W}^t ||^{1/t}$. Also, $\lim_{t \rightarrow \infty} \bm{W}^t = \bm{0}$ if and only if $\rho(\bm{W}) < 1$ \citep[][Theorem 5.6.12]{HJ1985}.  The spectral radius is therefore interpreted as a long term average of the maximum amplification away from the interior equilibrium per unit time $t$ following a perturbation. For use in the objective function of an optimisation, note that the spectral radius is a continuous function of a matrix $\bm{W}$. Not only is the spectral radius related to the matrix norm through the long term average amplification, but any matrix norm also provides an upper bound on the spectral radius, $\rho(\bm{W}) \leq || \bm{W} ||$. The matrix norm and spectral radius are thus closely linked. The spectral radius for a stable system will also be bounded between zero and one.

Here, both the competition matrix $\bm{A}$ and the long term average per unit time $t$ multiplication number (amplification) of the CDM model is optimised for the observed $\bm{\bar{x}}$. The interior equilibrium of the CDM model, if it exists, is unique by Eq. \eqref{eq:xeq}. The joint optimisation maximises competition, or niche overlap if these terms are considered synonymous, while minimising the spectral radius  $\rho(\bm{J}^\ast)$ that defines the long term average amplification of perturbations away from the interior equilibrium, $\bm{x}^\ast$.  If $\rho(\bm{J}^\ast) < 1$ then the deviations will decrease over time  such that the adult  abundances eventually return to  $\bm{x}^\ast$. Otherwise, the deviations may continue to increase and the species abundances will not return to the interior equilibrium. 

The entries of the Jacobian matrix for the CDM model depend only on the species equilibrium relative abundances and the given parameters  $\bm{z}$ and $\bm{\tilde{\lambda}}$ (Section \ref{sec:all}). Assume that the density independent fecundity and survival rates are given such that population growth will occur in the absence of competition, that is, $\bm{\tilde{\lambda}} > \bm{z}$.  Let the Jacobian of the CDM model as a function of species abundances $\bm{x}$ and competition matrix $\bm{A}$ be denoted by $\bm{J}(\bm{x}, \bm{A})$. For example, $\bm{J}^\ast = \bm{J}(\bm{x}^\ast, \bm{A})$ in Eq. \eqref{eq:Jeq} for the CDM model evaluated at the interior feasible equilibrium. By Eq. \eqref{eq:pG}, $\bm{J}(\bm{x}^\ast, \bm{A}) = \bm{J}(\bm{p}^\ast, \bm{A})$ where $\bm{p}^\star = \bm{x}^\star / \sum_{i \in \mathcal{M}} x_i^\star$. The following  minimisation problem is constructed:
\begin{align*}
	\bm{\hat{A}} = \arg\min_{\bm{A} \in \bm{S}_m} [\rho(\bm{J}(\bm{\bar{x}}, \bm{A})) - 1] \sum_{i, j = 1}^m A_{ij},
\end{align*}  
where $\bm{S}_m$ is the set of $m \times m$ candidate matrices for $\bm{A}$ and $\bm{\bar{x}}$ is the vector of estimated equilibrium abundances. 

The conjecture encapsulated by this objective function is that i) it will be favourable to species $i$ if it can outcompete its competitors for limited resources and so increase $A_{ij}$, $i \neq j$, and ii) species are more likely to persist and be observed near the interior equilibrium in a system with greater resilience and shorter return time. That is, a system with a larger spectral radius and hence larger average amplification of perturbations away from equilibrium will be less likely to persist and be observed near the equilibrium than one with a small spectral radius, where perturbations are relatively attenuated. The objective function also allows multiplication by a constant without effect on the optimisation. For example, multiplication by $m^{-2}$ allows interpretation of the summation in the objective function as a measure of the average competition strength defined by the mean of the entries of $\bm{A}$. 

Note that the choice of objective function optimises both increased competition and increased resilience of the equilibrium estimate $\bm{\bar{x}}$. The objective function does not simply seek to minimise the spectral radius at that point: The latter approach would be expected to favour diagonal $\bm{\hat{A}}$ because decreasing interspecific competition is stabilising as seen in Proposition \ref{prop:CDM}. For large systems and non-symmetric or unconstrained $\bm{S}_m$ the number of unknowns would increase at the rate $m^2$ and so outpace the information provided by the $m$-dimensional vector $\bm{\bar{x}}$. It is conceivable then that the objective function may not have a single global optimum. In other words, $\bm{\hat{A}}$ may not be identifiable. In such cases, it may be useful to include constraints on the candidate set of matrices $\bm{S}_m$ as informed by prior knowledge or theoretical justification.

A low dimensional numerical example is provided below. In this example, the candidate set for optimisation includes only the nonnegative matrices with main diagonal entries equal to one with reference to Remark \ref{remark:q}.  Lemma \ref{lemma:eq} then applies so that a positive $\bm{q}$ is guaranteed for any feasible interior equilibrium given any choice of matrix $\bm{B}$ in this candidate set. Furthermore, the canonical structure of $\bm{\alpha}$ described in Section \ref{sec:Pianka} is then adopted so that $\bm{S}_m$ is the set of symmetric nonnegative matrices with main diagonal entries equal to one.

\subsection{Example Application}\label{sec:app}

\citet{Pombi2017} apply Pianka's method (Section \ref{sec:Pianka}) that uses field estimates of relative abundances among species to estimate the niche overlap of three species of important malaria mosquito vectors within the \textit{Anopheles gambiae} sensu lato (s.l.) species complex: \textit{An. arabiensis} ($Aa$), \textit{An. coluzzii} ($Ac$) and \textit{An. gambiae} sensu stricto ($Ag$). For this analysis, the CDM model is applied so that $\bm{x}(t)$ corresponds to adult females of the above three species. Additionally, \citet{Pombi2017} provide the total observed abundances across three sites in Burkina Faso. These estimates are halved to account for an approximate 1:1 sex ratio, and are summarised here with the estimated niche overlaps from \citet{Pombi2017}:
\begin{equation*}
	\bm{\alpha} = \begin{bNiceMatrix}[first-row,first-col]
			& Aa & Ac & Ag  \\
		Aa & 1   	& 0.717   & 0.562  \\
		Ac & 0.717  & 1   & 0.492 \\
		Ag & 0.562  & 0.492   & 1 \\
	\end{bNiceMatrix}, \quad
	\bm{\bar{x}} = \begin{bNiceMatrix}[first-col]
		Aa & 907 \\
		Ac & 1123 \\
		Ag & 675
	\end{bNiceMatrix} \times \frac{1}{2}.   
\end{equation*}
For this example, it will be assumed that the population abundance estimate $\bm{\bar{x}}$ approximates the interior equilibrium  $\bm{x^\ast}$ of the CDM model (Section \ref{sec:optim}). The fecundity rates $\bm{\tilde{\lambda}}$, adult survival probabilities $\bm{\sigma}$ and delays $\bm{\delta}$ are assumed similar for all three species in the \textit{An. gambiae} s.l. complex using estimates provided by \citet[][]{North2018}. For $i \in \mathcal{M}$, the duration of the immature period is set to $\delta_i = 10$, maximum fecundity rate is $\tilde{\lambda}_i = 9/2$ per female per day (the oviposition rate estimate is halved since this CDM model only applies to females with assumed 1:1 sex ratio) and adult probability of mortality is $z_i = 0.125$ per day. Rates of hybridisation among the three species are assumed negligible. In this example CDM model, the survival probabilities are absorbed into the fecundity parameter $\bm{\tilde{\lambda}}$ so that $\sigma_{a_i \mid i} = 1$ for $a_i = 1, \ldots, \delta_i - 1$ with $\sigma_{\delta_i \mid i} = 1 - z_i$ for $i \in \mathcal{M}$. This specification corresponds to a latent lagged population of immatures. See \citet{Hosack2023} for an elaboration that uses the above estimate for $\bm{A} = \bm{\alpha}$ with  interspecific hybridisation in a risk analysis of a genetic biocontrol application.

\begin{figure}[ht]
	\begin{subfigure}{0.49\textwidth}
		\includegraphics[width=0.99\textwidth]{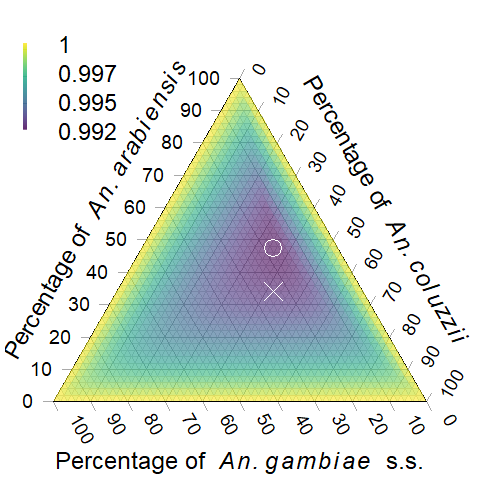}	
		\subcaption{Spectral radii given $\bm{A} = \bm{\alpha}$.}\label{fig:nicheTernary}
	\end{subfigure}
	\begin{subfigure}{0.5\textwidth}
		\includegraphics[width=0.99\textwidth]{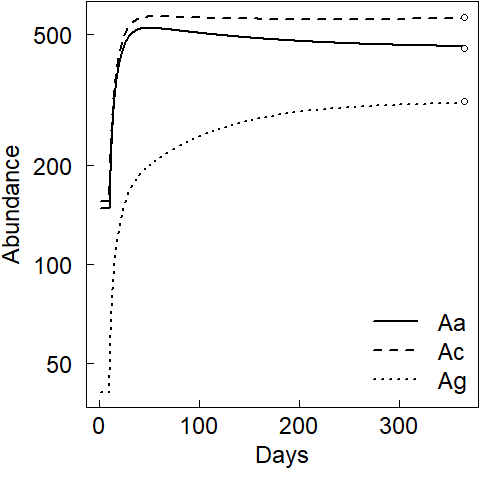}	
		\subcaption{Perturbation response given $\bm{A} = \bm{\alpha}$.}\label{fig:nicheTime}
	\end{subfigure}
	\begin{subfigure}{0.49\textwidth}
		\includegraphics[width=0.99\textwidth]{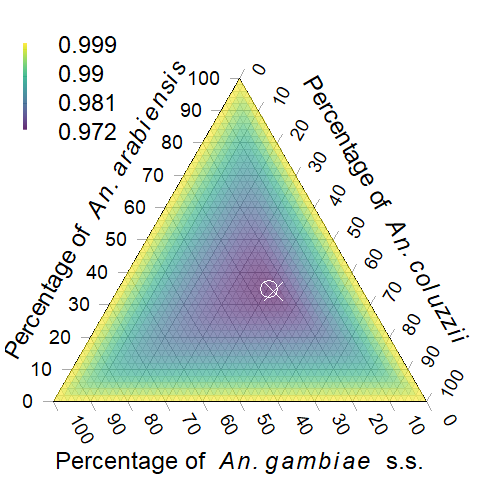}	
		\subcaption{Spectral radii given $\bm{A} = \bm{\hat{A}}$.}\label{fig:optTernary}
	\end{subfigure}
	\begin{subfigure}{0.5\textwidth}
		\includegraphics[width=0.99\textwidth]{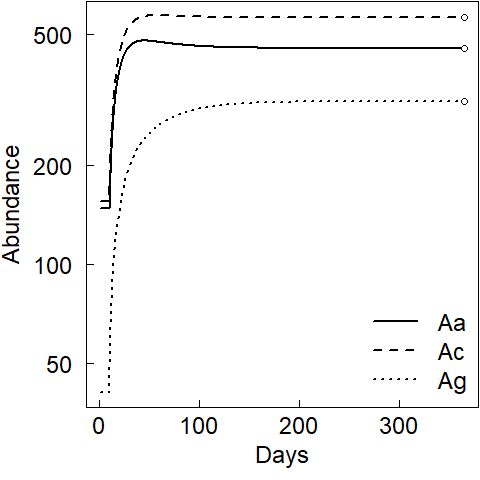}	
		\subcaption{Perturbation response given $\bm{A} = \bm{\hat{A}}$.}\label{fig:optTime}
	\end{subfigure}
	\caption{Spectral radius for all possible interior equilibrium species abundances given Pianka niche overlap estimate $\bm{A} = \bm{\alpha}$ (\textit{top}, a) and optimised estimate $\bm{A} = \bm{\hat{A}}$ (\textit{bottom}, c).  In (a) and (c), points show the location of equilibrium with lowest spectral radius ($\circ$) and the location of the abundance estimate $\bm{\bar{x}}$ ($\times$). Note change in colour legend of (a) versus (c) to aid visualisation. The return to equilibrium after a perturbation that decreases the species abundances by 33\%, 28\% and 13\% for the three species $Aa$, $Ac$ and $Ag$ is shown for $\bm{A} = \bm{\alpha}$ (\textit{top}, b) and $\bm{A} = \bm{\hat{A}}$ (\textit{bottom}, d). In (b) and (d), points at 365 days denote the equilibrium values of each species, $\bm{\bar{x}}$.}\label{fig:NicheOpt}
\end{figure}

Let $\bm{A} = \bm{\alpha}$ denote the assumption that the competition coefficients are identifiable with niche overlap indices estimated by the Pianka method, as suggested by \citet{May1975} and discussed in Section \ref{sec:Pianka}. The matrix 
$\bm{A} = \bm{\alpha}$ is nonnegative and nonsingular, and so the interior equilibrium is feasible by Lemma \ref{lemma:eq} with the knowledge that $\bm{\tilde{\lambda}} > \bm{z} > \bm{0}$ from the above estimates, and $\bm{q} > \bm{0}$ is then available from Eq. \eqref{eq:xeq}. The sufficient condition for local asymptotic stability of the interior equilibrium provided by Proposition \ref{prop:CDM} is not satisfied by $\bm{A} = \bm{\alpha}$. However, the numerical approach of Section \ref{sec:all} applies, and Figure \ref{fig:nicheTernary} shows that all numerically evaluated interior equilibria in the simplex, which correspond to different parameterisations of $\bm{q} \in \mathbb{R}_+^m$, are locally asymptotically stable for $\bm{A} = \bm{\alpha}$. The abundance estimate has corresponding relative abundance vector $\bm{p} = \bm{\bar{x}} / x_{tot} = [0.34, 0.42, 0.24]^\top$, where $x_{tot} = \sum_i \bar{x}_i$. This point does not correspond with the configuration of relative abundances that minimises the spectral radius. The spectral radius approaches one as the interior equilibrium approaches a boundary. Figure \ref{fig:nicheTime} shows the return trajectory to equilibrium after a random perturbation.

For comparison, the estimate $\bm{\hat{A}}$ is presented with the same abundance estimate $\bm{\bar{x}} \approx \bm{x^\ast}$ (and relative abundances $\bm{p}$) and proportional perturbations away from equilibrium. The set of possible competition matrices $\bm{S}_3$ is constrained to symmetric matrices with ones along the main diagonal for comparison with the niche overlap method. The resulting optimised estimate for competition is
\begin{equation*}
	\bm{\hat{A}} = \begin{bNiceMatrix}[first-row,first-col]
		& Aa & Ac & Ag  \\
		Aa & 1 & 0.34 & 0.25 \\ 
		Ac & 0.34 & 1 & 0.28 \\ 
		Ag & 0.25 & 0.28 & 1 \\ 
	\end{bNiceMatrix}.
\end{equation*}
The above estimates are roughly half the magnitude of the estimates provided by the niche overlap method, $\bm{\alpha}$, presented in \citet{Pombi2017}. The competitive system has an interior feasible equilibrium by Lemma \ref{lemma:eq} that is locally asymptotically stable by Proposition \ref{prop:CDM}. 

The basin of attraction for the optimised estimate $\bm{A} = \bm{\hat{A}}$ is deeper in Figure \ref{fig:optTernary} compared to the niche overlap estimate, and the minimum spectral radius is estimated to be near the abundance estimate. The optimisation decreases the spectral radius at $\bm{\bar{x}}$ while increasing the average strength of competition (Section \ref{sec:optim}). The optimisation thereby shifts the bottom of the ``bowl'' formed by the basin of attraction for the interior equilibrium $\bm{x}^\ast$ towards the estimate $\bm{\bar{x}}$.  This latter result, of course, may not always exactly correspond to $\bm{\bar{x}}$ given the constraints imposed on the set of possible candidate matrices $\bm{S}_3$ for fixed $\bm{\tilde{\lambda}}$ and $\bm{z}$. For $m = 3$ the number of unknowns given symmetric $\bm{A}$ with fixed main diagonal is the same as the number of species. The smaller spectral radius given optimised estimate $\bm{\hat{A}}$ corresponds to a faster return time to equilibrium following perturbation  (Figure \ref{fig:optTime}) compared to the niche overlap estimate (Figure \ref{fig:nicheTime}). The difference on the logarithmic scale is most apparent for \textit{An. arabiensis} and \textit{An. gambiae} s.s. that approach their equilibrium values noticeably more quickly in the optimised case. This optimal configuration  reflects faster rates of return of species abundances towards $\bm{\bar{x}}$ with reduced spectral radius in its linearised approximation compared to the niche overlap system, and so the perturbation is more dampened in the optimised system with greater resilience.

\section{Conclusion}

Sufficient conditions are presented for the local asymptotic stability of feasible interior equilibria within general multispecies systems of difference equations with delayed maturity.  Interspecific competition, mutualism, predation, and commensalism or amensalism are accommodated by the general model framework. The competitive delay multispecies (CDM) model is presented as an important specific example. A sufficient condition is provided for stability of coexistence in the CDM model that only requires stronger intraspecific competition relative to interspecific competition, regardless of delay. Moreover, evaluation of the simplex of relative abundances can exhaustively assess the stability properties of a system described by the CDM model given only competition $\bm{A}$, fecundity $\bm{\lambda}$ and survival probabilities $\bm{\sigma}$. In cases where competition $\bm{A}$ is unknown, a technique is provided to predict an optimal set of density dependence coefficients. A numerical illustration is provided for a low dimensional system, but future research would be needed to evaluate its suitability for very large systems. 

\begin{appendices}
	\section{Appendix}
	\subsection{Proof of Theorem \ref{prop:suff}}\label{app:suff}
\begin{proof}
	Each set $\mathcal{D}_1, \ldots, \mathcal{D}_r$ is formed by a unique set of $|\mathcal{D}_w| \geq 1$ species such that $\sum_{w} |\mathcal{D}_w| = |\mathcal{M}|$. Clearly, the spectrum $s(\bm{J}^\ast)$ with $\bm{J}^\ast$ in the triangular form of Eq. \eqref{eq:reducible} is the union of the eigenvalues of each   matrix $\bm{J}^\ast(\alpha_w, \alpha_w)$ with dimensions $(|\mathcal{D}_w| + \sum_{i \in \mathcal{D}_w} \delta_i) \times (|\mathcal{D}_w| + \sum_{i \in \mathcal{D}_w} \delta_i)$ that is irreducible or a zero entry for $w \in \{1, \ldots, r\}$. The condition for local asymptotic stability is then $\rho(\bm{J}^\ast(\alpha_w, \alpha_w)) < 1$ for $w \in \{1, \ldots, r\}$.
	
	If $|\mathcal{D}_w| = 1$ then $\mathcal{D}_w$ consists of a single node of species $i$ that does not form a cycle with any other node, and so $\delta_i = 0$. By Eq. \eqref{eq:Gpd0}, if 
	\begin{equation*}
		G_i(\bm{x}^\ast) + x_i^\ast  \left. \frac{\partial G_i(\bm{x}(t))}{\partial x_i(t)}\right|_{\bm{x}(t) = \bm{x}^\ast} < 1 - \sigma_{\delta_i \mid i} = z_i, \quad i \in \mathcal{D}_w, \quad |\mathcal{D}_w | = 1,
	\end{equation*}
	then $\rho(\bm{J}^\ast(\alpha_w, \alpha_w)) < 1$.
	
	If $|\mathcal{D}_w| > 1$ then each species $i \in \mathcal{D}_w$ has either age structure, $\delta_i > 0$, or a nontrivial cycle with at least one other species $j \in \mathcal{D}_w \backslash i$, or both. Let $\gamma_i'$ denote a cycle of minimum length that incorporates all age classes within species $i \in \mathcal{D}_w$. For $i \in \mathcal{D}_w$ with $|\mathcal{D}_w| > 1$, let $\Gamma_i$ denote the product of deleted absolute row sums that corresponds to $\gamma_i'$,
	\begin{align*}
		\Gamma_i &= \prod_{P_k \in \gamma_i'} R_k\left(\bm{J}^\ast(\alpha_w, \alpha_w)\right)\\ 
		&= \left\{\left|\left( G_i(\bm{x}^\ast) + x_i^\ast \left. \frac{\partial G_i(\bm{x}(t))}{\partial x_i(t)}\right|_{\bm{x}(t) = \bm{x}^\ast} \right) \right|  + \sum_{j \in \mathcal{D}_w\backslash i} \left| \left( \left. x_i^\ast\frac{\partial G_i(\bm{x}(t))}{\partial x_j(t)}\right|_{\bm{x}(t) = \bm{x}^\ast} \right)\right| \right\} \prod_{a = 0}^{\delta_i - 1} \sigma_{a \mid i}, \quad i \in \mathcal{D}_w.
	\end{align*}
	Among the rows of $\bm{J}^\ast(\alpha_w, \alpha_w)$ that contribute to the above $ R_k(\bm{J}^\ast(\alpha_w, \alpha_w)) $ of $\Gamma_i$, the only non-zero diagonal entry contributed by species $i$ is $\sigma_{\delta_i \mid i}$. The matrix $\bm{J}^\ast(\alpha_w, \alpha_w)$ is irreducible and hence weakly irreducible. Thus, a necessary condition for the bound on the spectral radius to be less than one is the condition
	\begin{equation*}
		\Gamma_i < 1 - \sigma_{\delta_i \mid i} = z_i, \quad \forall i \in \mathcal{D}_w,\label{eq:sigcon}
	\end{equation*}
	so that the maximum distance from the origin to the boundary of the eigenvalue region associated with $\gamma_i'$ by Theorem \ref{thm:Brualdi} is less than $\sigma_{\delta_i \mid i} + z_i = 1$.  The nodes included in cycle $\gamma_i'$ must also be included in any larger cycle that additionally incorporates any set $\mathcal{D}_w \backslash i$ of alternative species that communicate with species $i$. Moreover, each species $d \in \mathcal{D}_w \backslash i$ contributes the product $\Gamma_d$ associated with the minimal non-trivial cycle $\gamma_d'$ of that species, so that the overall product of absolute deleted row sums for any subset of species $\mathcal{E} \in \mathcal{D}_w$ is $\Gamma_{\mathcal{E}} = \prod_{i \in \mathcal{E}} \Gamma_i$. If the sufficient conditions $0 < \Gamma_i < z_i \leq 1$ are met for all $i \in \mathcal{D}_w$, then $\Gamma_{\mathcal{E}} < 1$ for any $\mathcal{E} \in \mathcal{D}_w$ and $\rho(\bm{J}^\ast(\alpha_w, \alpha_w)) < 1$. 
\end{proof}
	
	\subsection{Proof of Lemma \ref{lemma:Girr}}\label{app:Girr}
	\begin{proof}
		Let $D$ denote the directed graph of $\bm{J}^\ast$. For $\delta_i > 0$ a directed path in $D$ exists that includes all nodes of species $i$ by the delayed age structure of Eq. \eqref{eq:Gvec} because $\sigma_{a_i \mid i} > 0$ for $a_i = 0, \ldots, \delta_i - 1$. If the partial derivatives of Eq. \eqref{eq:Gpdx} and \eqref{eq:Gpd0} are non-zero for all cross partials of the zero age classes with the mature age classes, then Eq. \eqref{eq:Jeq} shows that a cycle in $D$ also exists that includes each node within a species to any other node, either within the species or in another species. $D$ is then strongly connected and $\bm{J}^\ast$ is irreducible.
	\end{proof}
	
	\subsection{Proof of Theorem \ref{prop:suff_alt}}\label{app:suff_alt}
	\begin{proof}
		The matrix $\bm{\hat{J}}$ may be permuted to have block structure
		\begin{equation*}
			\bm{\hat{J}} =
			\begin{bmatrix}
				\bm{{J}^\circ} & \bm{0} \\
				\ast & \bm{{J}^\bullet}
			\end{bmatrix}
		\end{equation*}
		that is conformally partitioned with respect to $\hat{\bm{x}} = [\bm{x}^\circ, \bm{x}^\bullet]^\top$ and with zero entries in the upper right quadrant from Eq. \eqref{eq:Jeq}, although not necessarily the lower left quadrant. The spectrum of eigenvalues $s(\bm{\hat{J}})$ is therefore the union of the spectra $s(\bm{{J}^\circ}) \cup s(\bm{{J}^\bullet})$. It is given that the maximum modulus of the eigenvalues in $s(\bm{{J}^\bullet})$ is less than one. Again by Eq. \eqref{eq:Jeq}, $\bm{{J}^\circ}$ is a block diagonal matrix with each of the $r$ blocks corresponding to a species $i$ with zero equilibrium abundance for the adult class, $x_i^\circ = 0$, $i = 1, \ldots, r$. The spectrum of eigenvalues for $\bm{{J}^\circ}$ is then the union of the spectra for each block. The largest cycle in each block is also the minimal cycle that includes all age classes. The condition of the proposition is sufficient for the moduli of eigenvalues to be less than one across the $r$ blocks by application of Theorem \ref{prop:suff} to each block.
	\end{proof}
	
	\subsection{Proof of Lemma \ref{lemma:eq}}\label{app:eq}
	\begin{proof}
		The interior equilibrium is unique by Eq. \eqref{eq:xeq}, and so it suffices to show that the solution for $\bm{q} = \textrm{diag}[\bm{z}]\textrm{diag}\left[\bm{\tilde{\lambda}} - \bm{z}\right]^{-1}\bm{A}\bm{x}^\ast$ is positive given $\bm{A}$, $\bm{\tilde{\lambda}}$, $\bm{z}$ and positive $\bm{x}^\ast > \bm{0}$. The $m \times m$ matrix $\textrm{diag}[\bm{z}]\textrm{diag}\left[\bm{\tilde{\lambda}} - \bm{z}\right]^{-1}$ is diagonal with positive entries along the main diagonal, and so $\bm{q} > \bm{0}$ if $\bm{A}\bm{x}^\ast > \bm{0}$. If $A_{ii} > 0$ for all $i \in \mathcal{M}$ then $\sum_{k = 1}^m A_{ik}x_k^\ast \geq A_{ii}x_i^\ast > 0$ for all $i \in \mathcal{M}$ since $\bm{x}^\ast > \bm{0}$ and $\bm{A} \geq \bm{0}$.
	\end{proof}
	
	\subsection{Proof of Proposition \ref{prop:CDM}}\label{app:CDM}
	\begin{proof}
		The constraint given the nonnegativity of $\bm{A}$ ensures that the main diagonal entries of $\bm{A}$ are positive, and so if the constraint holds then it is also true that $\bm{A}$ is diagonally dominant such that $| A_{ii} | > \sum_{j\in\mathcal{M}\backslash i} | A_{ij}| $ for all $i \in \mathcal{M}$. The latter constraint is a known result that ensures $\bm{A}$ is invertible \citep[see][Corollary 5.6.17]{HJ1985}. Lemma \ref{lemma:eq} then applies, and so also Eqs. \eqref{eq:xeq} and \eqref{eq:pG} hold for $\bm{q} > \bm{0}$ corresponding to $\bm{x}^\ast > \bm{0}$. Application of Theorem \ref{prop:suff} to the CDM model by Eqs. \eqref{eq:pdx} and \eqref{eq:pG} shows that the interior equilibrium $\bm{x}^\ast > \bm{0}$ is locally asymptotically stable if
		\begin{equation}
			z_i\frac{\lambda_i}{\tilde{\lambda}_i}\left(\left| 1 - \frac{\tilde{\lambda}_i - z_i}{\tilde{\lambda}_i}\frac{A_{ii} x_i^\ast}{\sum_{k\in\mathcal{M}} A_{ik}x_k^\ast} \right| + \sum_{j \in \mathcal{M}\backslash i} \frac{\tilde{\lambda}_i - z_i}{\tilde{\lambda}_i} \frac{A_{ij} x_i^\ast}{ \sum_{k \in \mathcal{M}} A_{ik} x^\ast_k} \right) \prod_{a = 0}^{\delta_i - 1} \sigma_{a \mid i} < z_i, \quad \forall i \in \mathcal{M}.\label{eq:propCDM}
		\end{equation}
		Given constraints $\bm{A} \geq \bm{0}$, $0 < z_i < 1$ and $\tilde{\lambda}_i - z_i > 0$, the difference within the absolute value in the above equation is always positive because $0 < \tilde{\lambda}_i - z_i < \tilde{\lambda}_i$ and $0 < A_{ii}x_i^\ast \leq \sum_{k \in \mathcal{M}} A_{ik} x^\ast_k$. By the definition $\tilde{\lambda}_i = \lambda \prod_{a = 0}^{\delta_i - 1} \sigma_{a \mid i}$, Eq. \eqref{eq:propCDM} is therefore equivalent to
		\begin{equation*}
			\frac{\tilde{\lambda}_i - z_i}{\tilde{\lambda}_i} \times \frac{x_i^\ast}{\sum_{k \in \mathcal{M}} A_{ik} x^\ast_k}\left[-A_{ii} + \sum_{j\in\mathcal{M} \backslash i} A_{ij}\right] < 0
			, \quad i \in \mathcal{M},\label{eq:iconstraint}
		\end{equation*}
		where the terms outside of the brackets on the left hand side are positive.
	\end{proof}
	
	\subsection{Proof of Corollary \ref{cor:CDMalt}}\label{app:CDMalt}
	\begin{proof}
		The result follows from Theorem \ref{prop:suff_alt} applied to the CDM model with Lemma \ref{lemma:eq} in force because Eq. \eqref{eq:G} has the property that $\hat{x}_i = x_i^\circ = 0 \implies \hat{x}_i G_i(\bm{\hat{x}}) = 0$ for $i \in \mathcal{M}$. The condition for local asymptotic stability from Theorem \ref{prop:suff_alt} is that $|G_i(\bm{\hat{x}})| \prod_{a = 0}^{\delta_i - 1}\sigma_{a \mid i} < 1$ for all $i \in \{j : \hat{x}_j = x_j^\circ\}$, where $\tilde{\lambda}_i = \lambda_i \prod_{a = 0}^{\delta_i - 1} \sigma_{a \mid i}$ and $G_i$ is defined by Eq. \eqref{eq:G}.
	\end{proof}
	
\end{appendices}

\section*{Acknowledgements}

The authors thank Lawrence Forbes and the anonymous reviewers for helpful comments. This work was supported, in whole or in part, by  a grant to the Foundation for the National Institutes of Health from the Bill \& Melinda Gates Foundation [INV-008525]. 

\section*{Data and Code Availability}

The reproducible code and simulated data is publicly available at:\\ \url{https://github.com/csiro-risk-assessment/MultiSppMaturityDelayStability}.

\bibliographystyle{chicago}
\bibliography{MultiSppCompetitionDelayedMaturation_refs}

\end{document}